\documentclass[10pt,conference]{NEMO}
\usepackage{graphicx}
\usepackage{multirow}
\usepackage{cite}
\usepackage{bm}
\usepackage{amsmath,amssymb,amsfonts}
\usepackage{algorithmic}
\usepackage{graphicx}
\usepackage{textcomp}
\usepackage{xcolor}
\usepackage{subcaption}
\usepackage{bbold}

\ifCLASSINFOpdf
\else
\fi

\makeatletter
\begin{document}
%
\title{Information Theoretic Analysis of a Dual-Band MIMO Cellphone Antenna with ANSYS HFSS SBR+}

\author{Volodymyr~Shyianov,~Bamelak~Tadele,~Vladimir~I.~Okhmatovski,~and~Amine~Mezghani\\
		\\
University~of~Manitoba,~Winnipeg,~Manitoba,~R3T~5V6,~Canada}




\maketitle

\begin{abstract}
Historically, the design of antenna arrays has evolved separately from Shannon theory. Shannon theory adopts a probabilistic approach in the design of communication systems, while antenna design approaches have relied on the deterministic Maxwell theory alone. In this paper, we investigate an information-theoretic analysis approach which we apply to evaluate the design of a dual-band, dual-polarized  multiple-input multiple-output (MIMO) array on a cellphone. To this end, we use ANSYS HFSS, a commercial electromagnetic (EM) simulation software suitable for the numerical optimization of antenna systems. HFSS is used to obtain an accurate model of the cellphone MIMO antenna array and HFSS SBR+ is utilized to obtain channel matrices for a large number of users. Taking advantage of linear and optimal processing at the cellphone, we estimate the outage probability curves. The curves are then used to determine the diversity gain in a moderate signal-to-noise ratio (SNR) regime and the multiplexing gain at a high SNR regime. This approach is then compared with the method of estimating the diversity gain from the envelope correlation coefficients or the beam-coupling matrix showing substantial differences in the two methodologies.      
\end{abstract}


%
\IEEEpeerreviewmaketitle

\section{Introduction}
 Consider a transmitter equipped with $N_t$ antenna elements which aims to communicate with a receiver that is endowed with $N_r$ antenna elements. The mathematical representation of the underlying MIMO channel is given by:
\begin{equation}\label{eq:MIMO_channel}
    \mathbf{y} = \mathbf{H}\mathbf{x}+\mathbf{n},
\end{equation}
where $\mathbf{n}$ represents the additive noise. This widely adopted noisy linear model has led to great developments from both the perspective of signal processing and information theory \cite{tse2005fundamentals,heath2018foundations}. While this model captures the essence of Maxwell's equations, the precise description of the channel matrix $\mathbf{H}$ and the statistics of the noise $\mathbf{n}$ remains an active research topic. Multi-port communication theory, first introduced in \cite{wallace2004mutual} and popularized by \cite{ivrlavc2010toward}, offers a consistent approach to incorporate the physics of radio-communication into the model of the channel matrix and the noise statistics. Based on the model (\ref{eq:MIMO_channel}) we can use multiport communication theory \cite{ivrlavc2010toward} to represent the input signal, $\mathbf{x}$, by transmit voltage sources, and the output signal, $\mathbf{y}$, by the voltages received at the load. Under this paradigm, the channel matrix $\mathbf{H}$, at any frequency, can be found from cascaded multiport networks and is given by \cite{tadele2023channel}: 
\begin{eqnarray}\label{eq:Scattering_Channel_MIMO}
     \mathbf{H}(f)  = \big[\mathbf{I}+\mathbf{S}_{\mathrm{L}}(f)\big]\big[\mathbf{I} - \mathbf{S}_{\mathrm{R}}(f)\mathbf{S}_{\mathrm{L}}(f)\big]^{-1}
     &\\ \nonumber& \!\!\!\!\!\!\!\!\!\!\!\!\!\!\!\!\!\!\!\!\!\!\!\!\!\!\!\!\!\!\!\!\!\!\!\!\!\!\!\!\!\!\!\!\!\!\!\!\!\!\!\!\!\!\!\!\!\!\!\!\!\!\!\!\!\!\!\!\!\!\!\!\!\!\!\!\!\!\!\!\!\!\!\!\!\!\!\!\!\!\!\!
    \times ~
    \mathbf{S}_{\mathrm{RT}}(f)\big[\mathbf{I}-\mathbf{S}_{\mathrm{S}}(f)\mathbf{S}_{\mathrm{T}}(f)\big]^{-1}\big[\mathbf{I}-\mathbf{S}_{\mathrm{S}}(f)\big].
\end{eqnarray}
Here the transmit antenna scattering matrix is given by $\mathbf{S}_{\mathrm{T}}(f)$, the receive antenna scattering matrix is given by $\mathbf{S}_{\mathrm{R}}(f)$, and the propagation medium will be modeled through the use of $\mathbf{S}_{\mathrm{RT}}(f)$. In the model, the linear generator with the associated matching network is represented through the source scattering matrix $\mathbf{S}_{\mathrm{S}}(f)$. The low-noise amplifier with the associated matching network is modeled through the load scattering matrix $\mathbf{S}_{\mathrm{L}}(f)$. For the remainder of this paper we assume that source and load terminations are $50 ~[\Omega]$ such that $\mathbf{S}_{\mathrm{S}}$ = $\mathbf{0}$ and $\mathbf{S}_{\mathrm{L}}$ = $\mathbf{0}$. 
\section{HFSS SBR+ Model}
\begin{figure}[h!]
    \centering
    \includegraphics[width=1\linewidth]{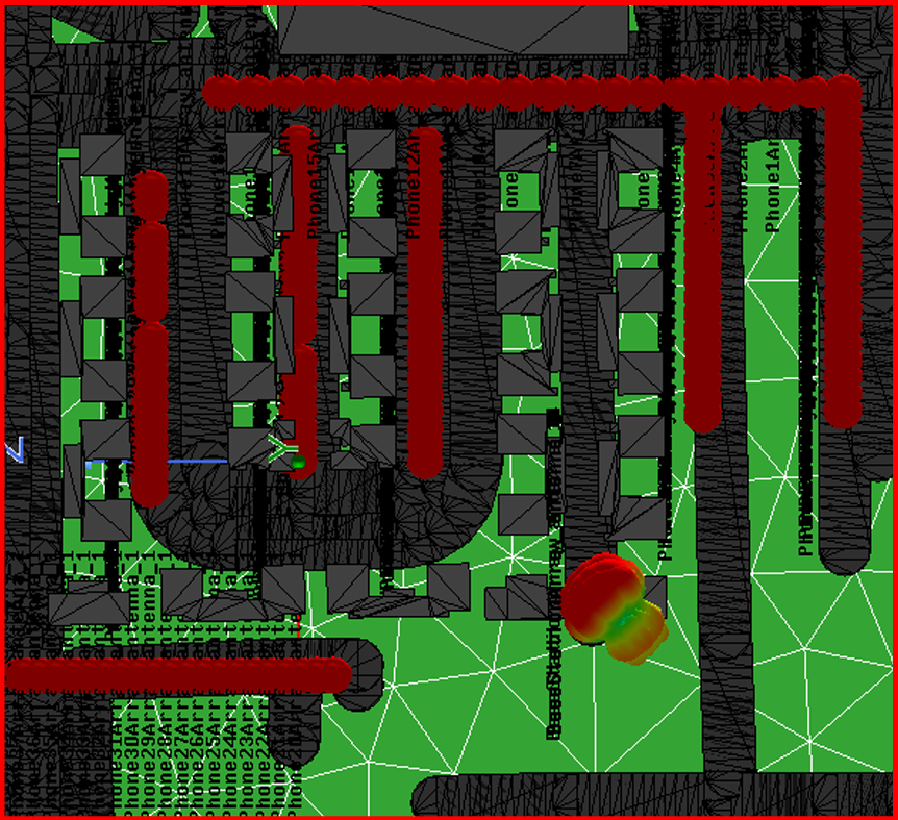}
    \vskip 0.1cm
    \caption{Cupertino 140 Users, single base-station array} 
    \label{fig:Cupertino}
\end{figure}

In Fig. \ref{fig:Cupertino} we show the imported open-street maps (OSM) geometry of a section of Cupertino, California chosen for analysis. The coordinates used for the import were ($37.3230 \textrm{N}^{\textrm{o}}$,  $122.0322 \textrm{W}^{\textrm{o}}$). The geometry consists of buildings, roads, and terrain, each configured as a single layer impedance with infinite thickness. The material properties used were those of concrete with relative permittivity of $\epsilon_r\approx 10$ and bulk conductivity of $\sigma \approx 1.7e^{-5}~[\textrm{S}/\textrm{m}]$.
A total of 140 user equipments (UEs) were distributed on the roads at a close proximity of $90-200~[\textrm{m}]$ away from the base-station.

\section{HFSS Antenna Model}
In Fig. \ref{fig:Cellphone} we show parts of the assembly of the dual-band cellphone UE antenna model implemented in Ansys HFSS. The assembly consists of an epoxy substrate FR4 backed by a perfect electric conductor (PEC) on which the antennas are mounted, a battery, and a glass screen. The substrate FR4 epoxy dielectric has relative permittivity of $\epsilon_r \approx 4.4$ and dielectric loss tangent $\delta\approx0.02$. The glass for the screen is lossless with relative permittivity $\epsilon_r = 5.5$. The battery is assumed to be made of PEC. The two cellphone antennas in Fig. \ref{fig:Cellphone} are located at the top left and bottom right corners of the assembly. The antenna on the top left of the assembly is designed for $2.46~[\textrm{GHz}]$ while the antenna in bottom right corner is designed for $3.16~[\textrm{GHz}]$. At their respective frequencies of operation both of the antennas are almost perfectly matched each with $S_{11}\approx -20~[\textrm{dB}]$. At $3.16~[\textrm{GHz}]$, the antenna at the top left corner of the assembly is mismatched, with $|S_{11}|\approx 0.7860$. This power mismatch between  the two antennas represents the difficulty of the dual-band operation where the degrees of freedom in both frequency and space are to be utilized. {In such instances a dual-band matching network needs to be optimized through information-theory \cite{shyianov2021achievable,saab2021optimizing}, since  greater tolerance of match decreases the overall system bandwidth \cite{Fano}}.  
\begin{figure}[h!]
    \centering
    \includegraphics[width=0.5\linewidth]{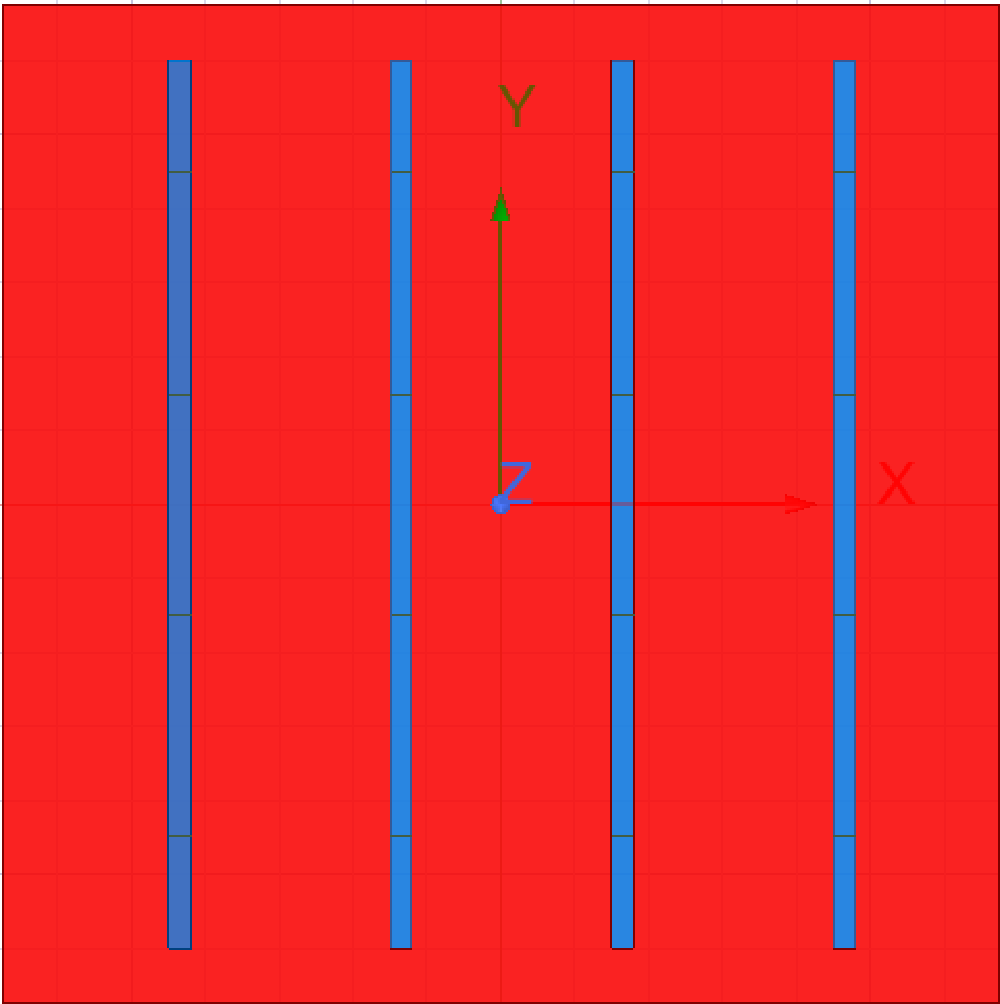}
    \vskip 0.1cm
    \caption{Planar slot base-station array with 16 antenna elements } 
    \label{fig:Cellphone}
\end{figure}
For the base-station array, we utilize a planar connected ultra-wideband slot array with 16 antenna elements designed to operate between $1$-$5~[\textrm{GHz}]$, where $5~[\textrm{GHz}]$ is the largest frequency of operation. Slot array antennas have previously been investigated in the context of connected arrays \cite{cavallo2011connected}. Such arrays have demonstrated large, theoretically infinite, bandwidth with the increase of the array size. As such, the connected slot structure is a promising design for massive MIMO base station arrays \cite{akrout2021achievable}.   
\begin{figure}[h!]
    \centering
    \includegraphics[width=0.5\linewidth]{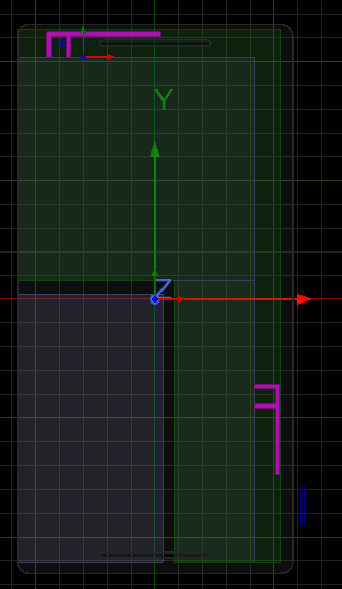}
    \vskip 0.1cm
    \caption{Cellphone MIMO antenna } 
    \label{fig:Cellphone}
\end{figure}
\section{Information Theoretic Analysis}\label{Sec: Info Theory}
 In this section we describe the chosen transmit and receieve processing schemes (both linear and optimal). In addition, we demonstrate the associated throughput calculations that will be utilized in section \ref{Sec:V} to determine the multiplexing and diversity gains from using two orthogonally polarized antennas on the cellphone.
\subsection{One Layer 1x1}
The baseline case in our study is the performance achieved by a single-layer, multiple-input single-output (MISO), system. In this setting, the signal of the mismatched antenna is ignored and only the dedicated antenna signal, designed for its band of operation, is used for data communication. The base-station is performing maximal-ratio transmission (MRT) towards a single cellphone antenna at either $3.16~[\textrm{GHz}]$ or $2.46~[\textrm{GHz}]$. The signal model at the receiver is given by,
\begin{equation}\label{eq:one_layer_11}
    y = \mathbf{h}^{\textrm{T}}\mathbf{w}x+n,
\end{equation}
where for MRT $\mathbf{w} = \frac{\mathbf{h^{*}}}{||\mathbf{h}||}$. With codebook vectors chosen from complex normal distribution $x\sim\mathcal{CN}(0,P_x)$ where $P_x$ is the time average signal power and additive white gaussian noise (AWGN) $n\sim\mathcal{CN}(0,P_n)$ with time average noise power $P_n$, the achievable throughput is determined from 
\begin{equation}\label{eq:MISO Rate}
    R = \mathrm{B}\log_2\left(1 + \frac{||\mathbf{h}||^2P_x}{P_n}\right) [\textrm{bits/s}],
\end{equation}
where $\mathrm{B}~[\textrm{Hz}]$ is the bandwidth in a single band.
\subsection{One Layer 2x1}
By combining the received signal at the two cellphone antennas, an improvement in throughput can be achieved. In this scenario, the base-station performs MRT towards either of the two antennas where thereafter the received signal is combined using maximal-ratio combining (MRC) at the cellphone. The signal model at the receiver (per antenna) is then given by:
\begin{equation}\label{eq:one_layer_21}
    y_1 = \mathbf{h}_1^{\textrm{T}}\mathbf{w}x+n_1,
\end{equation}
\begin{equation}
    y_2 = \mathbf{h}_2^{\textrm{T}}\mathbf{w}x+n_2,
\end{equation}
where for MRT $\mathbf{w} = \frac{\mathbf{h}_1^{*}}{||\mathbf{h}_1||}$. Assuming $n_1$ and $n_2$ are i.i.d complex normal random variables with time average power $P_n$, the achievable throughput is determined from,
\begin{equation}\label{eq:MISO Rate}
    R = \mathrm{B}\log_2\left(1 + \frac{(||\mathbf{h}_1^{\textrm{T}}\mathbf{w}||^2+||\mathbf{h}_2^{\textrm{T}}\mathbf{w}||^2)P_x}{P_n}\right) ~[\textrm{bits/s}].
\end{equation}
\subsection{MIMO Linear Processing}
By beamforming towards both of the cellphone antennas using MRT and applying linear minimum mean-squared error (LMMSE) equalization at the cellphone, we get the following system model,
\begin{equation}\label{eq:MIMO_linear}
    \mathbf{y} = \mathbf{L}\mathbf{H}\mathbf{W}\mathbf{x}+\mathbf{L}\mathbf{n},
\end{equation}
where $\mathbf{W} = \frac{\mathbf{H}^{\textrm{H}}}{||\mathbf{H}||}$ and $\mathbf{L}$ is LMMSE equalizing matrix. From (\ref{eq:MIMO_linear}) we obtain the equivalent signal model,
\begin{equation}\label{eq:MIMO_equivalent_1}
    y_1 = \widetilde{{h}}_1x_1+e_1,
\end{equation}
\begin{equation}
    y_2 = \widetilde{{h}}_2x_2+e_2,
\end{equation}\label{eq:MIMO_equivalent_2}
where $\widetilde{h}_1$, $\widetilde{h}_2$ are scalar equivalent channels after equalization and $e_1$, $e_2$ combine the effects of noise and equalization error. In terms of achievable throughput, we distinguish the cases of one layer and two layer systems. For single layer transmission, the throughput is given by
\begin{equation}\label{eq:MIMO Rate One Layer}
    R = \mathrm{B}\log_2\left(1 + \frac{(|\widetilde{h}_1|^2+|\widetilde{h}_2|^2)^2P_x}{|\widetilde{h}_1|^2P_{e_1}+|\widetilde{h}_2|P_{e_2}}\right)~ [\textrm{bits/s}].
\end{equation}
For two layer transmission, the throughput is given by
\begin{equation}\label{eq:MIMO Rate Two Layer}
    R = \mathrm{B}\log_2\left(1 + \frac{|\widetilde{h}_1|^2P_x}{P_{e_1}}\right) + \mathrm{B}\log_2\left(1 + \frac{|\widetilde{h}_2|^2P_x}{P_{e_2}}\right)~[\textrm{bits/s}].
\end{equation}
\subsection{MIMO Optimal Processing}
With optimal processing at the cellphone, such as MMSE-SIC receiver architecture, we achieve a MIMO mutual information bound given by
\begin{equation}\label{eq:MIMO Rate Two Layer}
    R = \mathrm{B}\log_2\det\left(\mathbf{I} + \mathbf{H}^{\textrm{H}}\mathbf{W}\mathbf{W}^{\textrm{H}}\mathbf{H} \right) ~[\textrm{bits/s}],
\end{equation}
where $\mathbf{W} = \frac{\mathbf{H}^{\textrm{H}}}{||\mathbf{H}||}$ and $\mathbf{I}$ is the identity matrix.
\section{Simulation Results and Discussion}\label{Sec:V}
\begin{figure}[h!]
    \centering
    \includegraphics[width=1\linewidth]{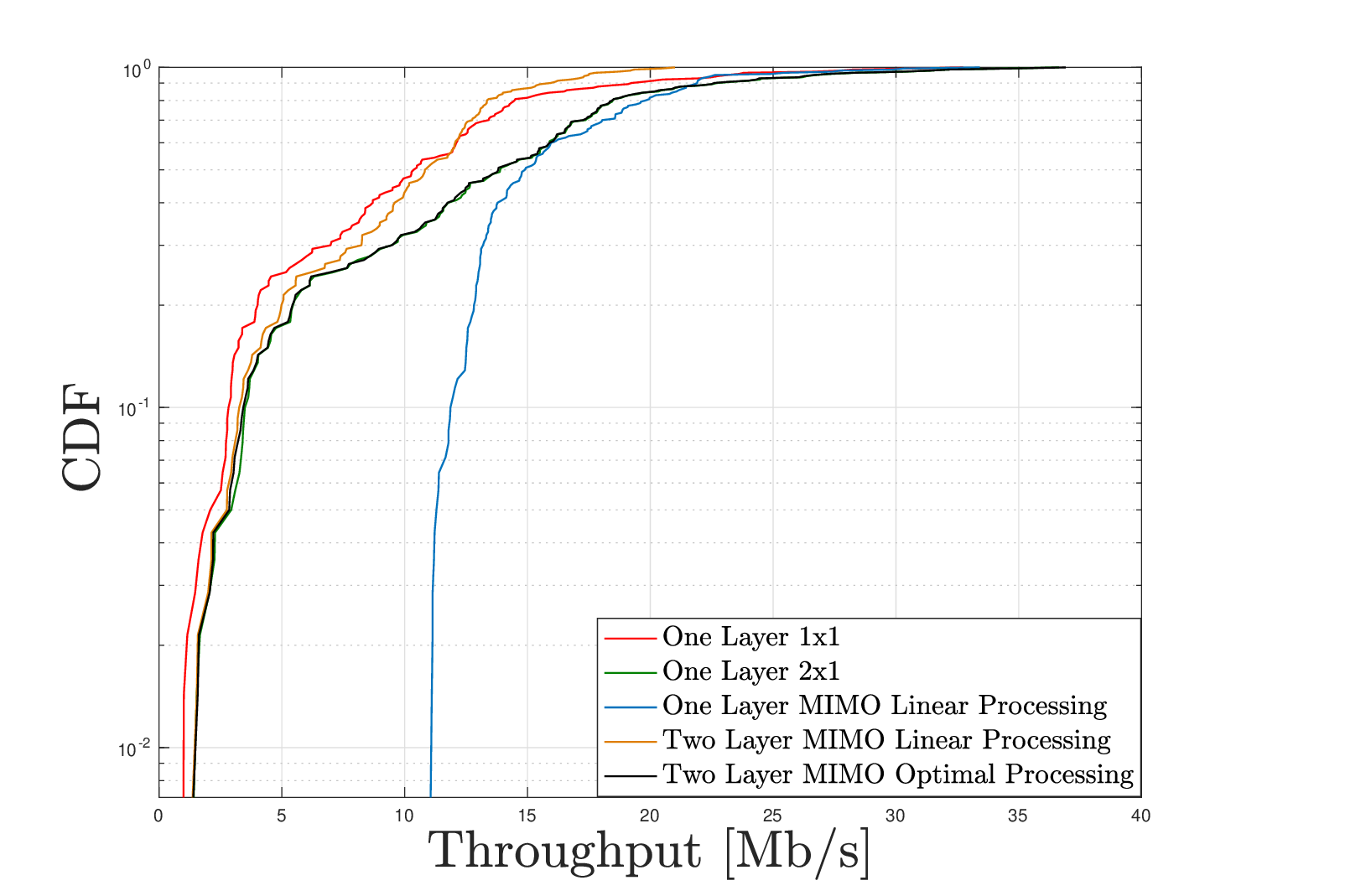}
    \vskip 0.1cm
    \caption{CDF of throughput at noise power of $-60~[\textrm{dBm}]$.} 
    \label{fig:Low SNR}
\end{figure}

\begin{figure}[h!]
    \centering
    \includegraphics[width=1\linewidth]{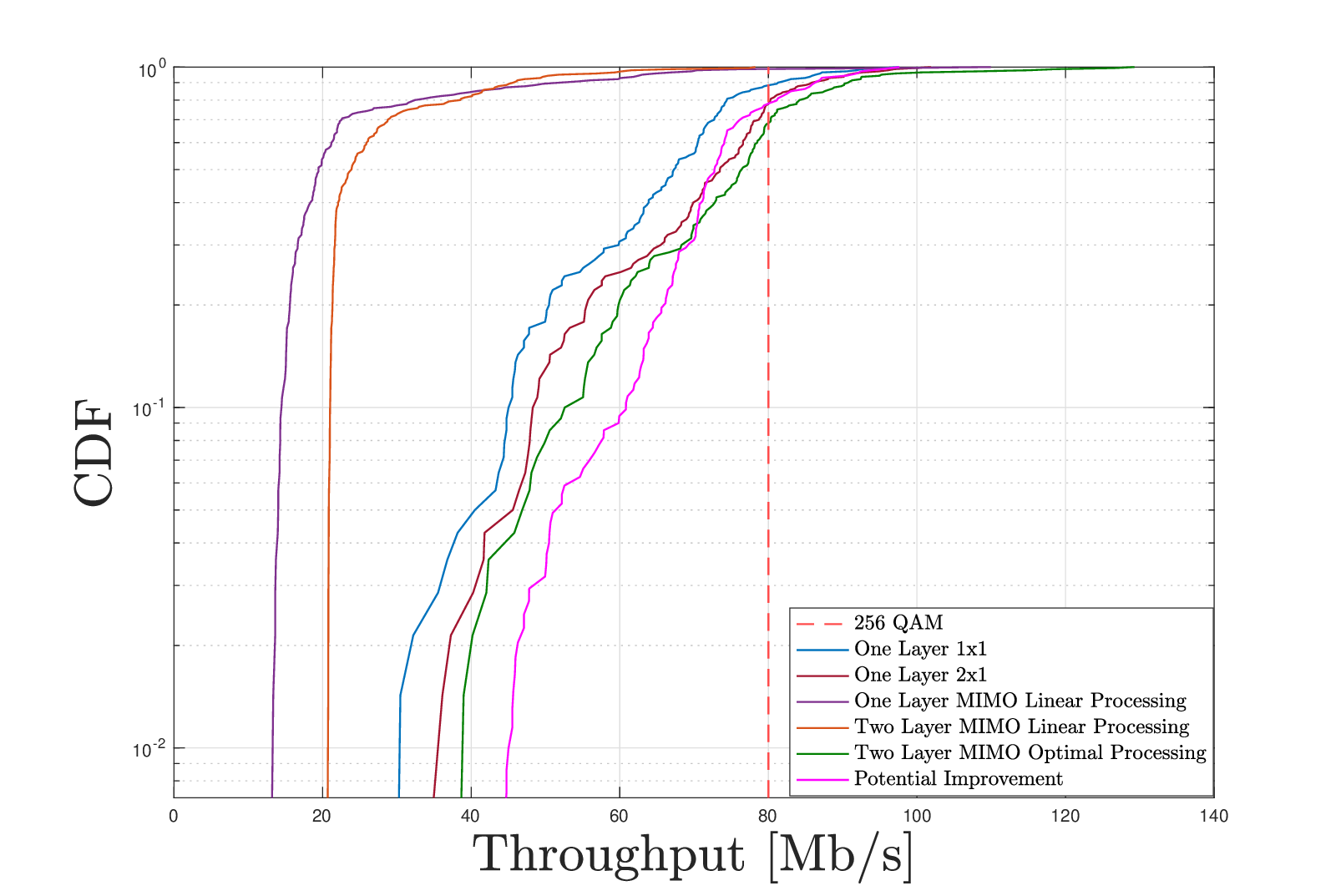}
    \vskip 0.1cm
    \caption{CDF of throughput at noise power of $-80~[\textrm{dBm}]$.} 
    \label{fig:Mid SNR}
\end{figure}

\begin{figure}[h!]
    \centering
    \includegraphics[width=1\linewidth]{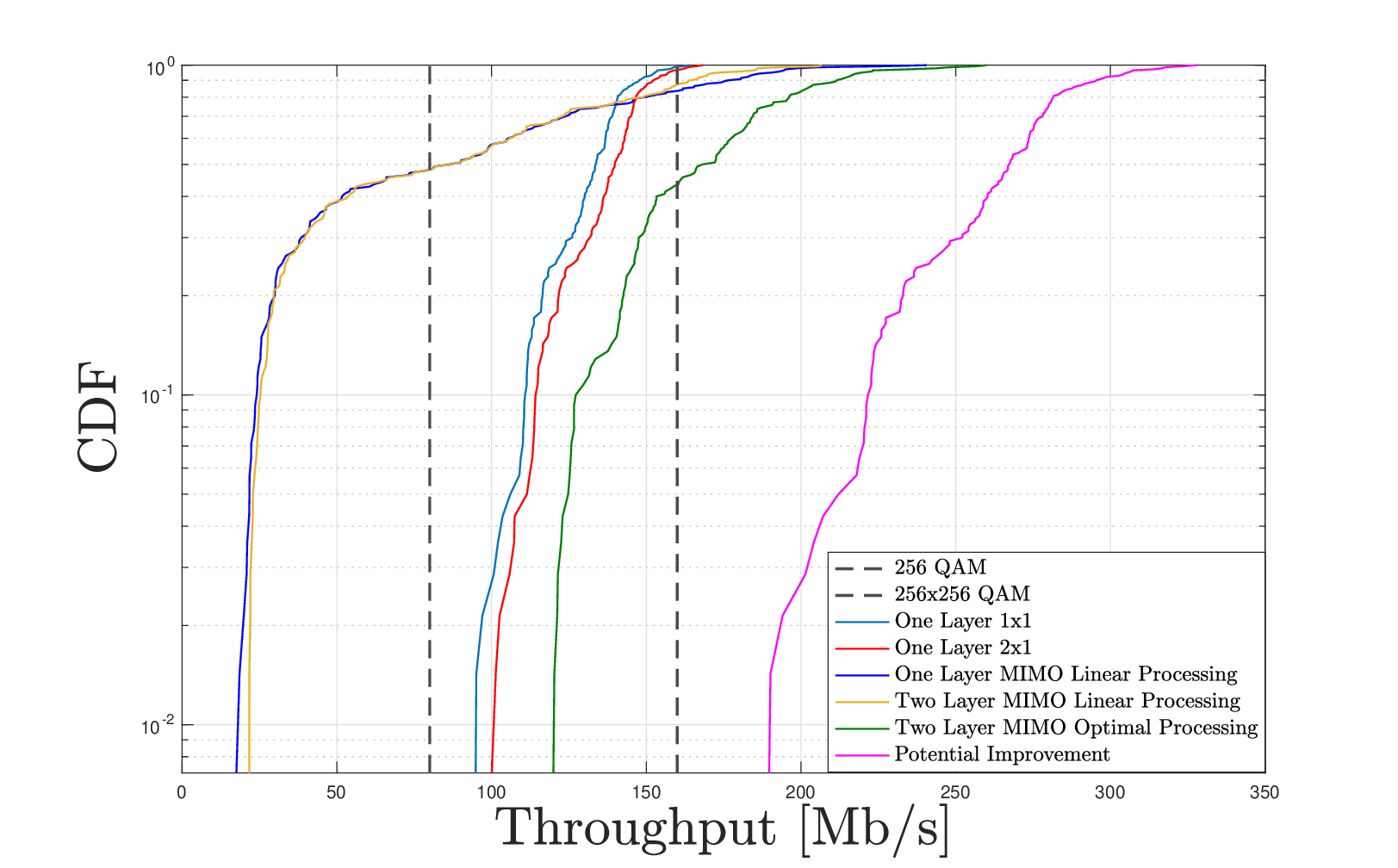}
    \vskip 0.1cm
    \caption{CDF of throughput at noise power of $-100~[\textrm{dBm}]$.} 
    \label{fig:High SNR}
\end{figure}

In this section we discuss the results of the antenna analysis based on the estimated outage probability curves for the transmit and receive processing schemes described in section \ref{Sec: Info Theory}. The entire analysis is performed at $3.16~[\textrm{GHz}]$ using a bandwidth of $10~[\textrm{MHz}]$. In Fig. \ref{fig:Low SNR}, the low \textrm{SNR} regime, the time average noise power is set to $P_n = -60~[\textrm{dBm}]$ which makes the average receive $\textrm{SNR} \approx -10~ [\textrm{dB}]$. In Fig. \ref{fig:Mid SNR}, the middle \textrm{SNR} regime, the time average noise power is set to $P_n = -80~[\textrm{dBm}]$ which makes the average receive $\textrm{SNR} \approx 10 ~[\textrm{dB}]$. In Fig. \ref{fig:High SNR}, the high \textrm{SNR} regime, the time average noise power is set to $P_n = -100~[\textrm{dBm}]$ corresponding to an average receive $\textrm{SNR} \approx 30 ~[\textrm{dB}]$.
From Fig. \ref{fig:Low SNR}, we note that it is not advantageous to use multiple data streams at such low levels of \textrm{SNR}. Both linear and optimal MIMO processing at the cellphone fail to improve the data rate over a simple receive diversity combining scheme. This result stems from the fact that at $\textrm{SNR} \approx -10~[\textrm{dB}]$, there is insufficient power to split between the two layers. However, with a single layer MIMO, much improvement can be achieved as can be seen in the blue curve. The reason for such an improvement comes from base-station processing rather than UE processing. With the base-station creating two beams, which can experience independent fading, we are able to take advantage of the transmit diversity and avoid  deep fades.  
In Fig. \ref{fig:Mid SNR}, we demonstrate the importance of multi-antenna diversity at the cellphone. Comparing the blue curve with the maroon curve we observe a diversity gain of $d\approx 1.4$, where the diversity gain is defined as:
\begin{equation}
    d ~=~ \frac{\log_{10}(p^\textrm{2x1}_{\textrm{outage}})}{\log_{10}(p^\textrm{1x1}_\textrm{outage})}.
\end{equation}
The optimal diversity gain of $d=2$ is plotted in pink and shows that a potential throughput improvement of approximately $10 ~[\textrm{Mb/s}]$ or $20\%$ could be achieved with a better antenna design. Both linear MIMO schemes demonstrate weak performance due to the strong correlation in the channel matrix. In these cases, the linear schemes are interference-limited and have far worse performance than the optimal MMSE-SIC receiver shown in green. However, optimal MIMO processing at the cellphone does not lead to sufficient improvement with respect to simple diversity combining to consider MIMO for the middle SNR regime. The calculated rates are also available with 256-QAM modulation supported by 3GPP 5G NR standards. This makes diversity a proper metric for optimization in this middle SNR regime.
Finally, we demonstrate the importance of MIMO multiplexing in the high SNR regime in Fig. \ref{fig:High SNR}. By comparing the blue and red curves, we notice that the diversity gain stays constant (compared to Fig. \ref{fig:Mid SNR}) yet the multiplexing gain, blue to green curves, becomes more substantial. The optimal multiplexing gain of $2$ is shown in pink which again demonstrates the potential improvements that can be obtained from careful antenna design. In this SNR regime, we see a more notable gain from multiplexing rather than diversity thereby making the multiplexing gain the metric we would want to assess our antenna designs on in this high SNR regime.

Looking at conventional metrics used for designing MIMO antennas such as isolation and envelope correlation coefficient (ECC) we find that the cellphone MIMO antenna has an $|S_{21}|=-22~[\textrm{dB}]$ and $\textrm{ECC}=0.04$. These metrics demonstrate a high level of isolation and polarization purity. With more advanced approaches based on the beam-coupling matrix \cite{kshetrimayum2022diversity} the calculated diversity gain for our cellphone MIMO antenna is approximately 2, which is the maximum diversity obtainable from two antennas and has not been verified by our analysis which has shown a mere diversity gain of 1.4. The approach in \cite{ivrlac2003quantifying} which relies on an estimate of the correlation matrix is more thorough yet suffers from the problem of power imbalance of the two antennas making the soft-rank approach they take a poor measure of diversity.

\section{Conclusion}
 In this paper, we investigated an information-theoretic approach to evaluate the design of a dual-band and dual-polarized cellphone MIMO array. To this end, we used HFSS to obtain an accurate model of the cellphone MIMO antenna array and HFSS SBR+ was utilized to obtain channel matrices for a large number of users in a complex scattering environment. Taking advantage of linear and optimal processing at the cellphone, we made estimates of the outage probability curves. The curves were then used to determine the diversity gain in a middle SNR regime, and multiplexing gain at a high SNR regime of the cellphone MIMO array. The approach was then compared to the traditional process of estimating the diversity gain from the envelope correlation coefficient and the beam-coupling matrix showing substantial differences in the two approaches and the value of doing a more thorough analysis as we have demonstrated.

\bibliographystyle{IEEEtran}
\bibliography{IEEEabrv,references}

\end{document}